\documentclass[12pt]{iopart}
\usepackage{iopams}  

\usepackage{graphics} 
\usepackage{graphicx}  
\usepackage{amsfonts}  
\usepackage{amssymb}
\usepackage{mathrsfs}
\usepackage{subfigure}
\usepackage{color}

\renewcommand{\phi}{\varphi}
\renewcommand{\>}{\right \rangle}

\newcommand{\ket}[1]{\left |#1\>}

\newcommand{\be}{\begin{equation}}
\newcommand{\ee}{\end{equation}}
\newcommand{\bea}{\begin{eqnarray}}
\newcommand{\eea}{\end{eqnarray}}

\begin{document}

\title[Multiple Ionization of Neon under soft x-rays: Theory vs Experiment]{Multiple Ionization of Neon under soft x-rays: Theory vs Experiment}

\author{G M Nikolopoulos$^1$ and P Lambropoulos$^{1,2}$}

\address{$^1$Institute of Electronic Structure \& Laser, FORTH, P.O.Box 1527, GR-71110 Heraklion, Greece}
\address{$^2$Department of Physics, University of Crete, P.O. Box 2208, GR-71003 Heraklion, Crete, Greece}


\begin{abstract}

We present a rather elaborate theoretical model describing the dynamics of Neon under radiation of photon energies $\sim 93$ eV and pulse duration in the range of 15 fs, within the framework of
Lowest non-vanishing Order of Perturbation Theory (LOPT), cast in terms of rate equations. Our model includes sequential as well as direct multiple ionization channels from the 2s and 2p atomic shells, including aspects of fine structure,  whereas the stochastic nature of SASE-FEL light pulses is also taken into account.  
Our predictions for the ionization yields  of  the different ionic species are in excellent agreement with the related experimental observations at FLASH. 

\end{abstract}

\maketitle


\section{Introduction}
\label{sec1}

In a 2011 paper \cite{LamPRA11}, we explored the conditions under which direct multiple ionization
channels might make significant contribution to ionic yields, which are usually dominated by the
inevitably present sequential channels. In order to present a quantitative assessment, we had chosen the
Neon atom driven by radiation of photon energy 93 eV, of pulse intensity and duration available at present
day  Free-Electron Laser (FEL) facilities such as FLASH \cite{Ack07,richter}. Our choice was motivated in
part by related experimental data obtained under the above mentioned conditions \cite{richter,richter09}. 
Since the chief objective of our study was the evaluation of direct multiple ionization in comparison to
the sequential contributions, we focused our model on the ionization of outer subshell (2p) electrons,
although a complete theoretical description would have required the inclusion of single-photon ionization
of 2s electrons as well. This means that we were evaluating the relative importance
of direct multiple ionization from the 2p subshell alone. In a sense, our work could be viewed as a
numerical experiment. At the time of that work, we were aware of only TOF (Time of Flight) data
\cite{richter09} which we did consider in the spirit of a qualitative comparison with our
calculations; because extracting ionic yields from a figure showing TOF spikes only is highly problematic.
Given its limited scope, that comparison was nevertheless compatible with our results, in the sense that
under the parameters of that experiment (in particular the pulse duration),  we did not expect a
discernible presence of contributions from the direct channels; and indeed none was found in the experimental 
data. It did, however, transpire that direct multielectron channels can begin competing with the sequential ones only
 when the duration of the pulse falls below 5 fs or so.

In a most recent paper by Guichard et al. \cite{Guichard13} addressing the same problem, 
the authors have presented a quantitative interpretation of the experimental data pertaining to the above
mentioned experiment, including this time laser intensity dependences of the ionic yields. This new piece
of experimental evidence behoves us to test our approach in a suitably more elaborate model. The approach
in Ref. \cite{Guichard13} relies on a rather simplified model, referred to by the authors as
``minimal model", in terms of which a good fit to the experimental data was obtained. That fit was compared
(see Fig.3 of Ref. \cite{Guichard13}) to what our equations would have given. Not surprisingly, our
equations lead to ionic yields systematically lower than the experimental data. That was to be expected
since, for the reasons outlined above, the single-photon ionization channels of the 2s electrons had not
been taken into account in \cite{LamPRA11}. Moreover, in \cite{Guichard13}  the authors compared the
experimental data to our results of Ref. \cite{LamPRA11}, for a Fourier-limited pulse duration of 30 fs,  
which however  is considerably  longer than the estimated pulse duration in the experiment i.e., $15\pm 5$
fs.
 
Since the term ``minimal model" may be open to a variety of interpretations, it is important to state  and
discuss here clearly the main assumptions underlying the model of Ref. \cite{Guichard13}. The formalism
rests on a set of rate equations governing the production and depletion of the ionic yields during the
pulse. For the photon energy, range of intensities and pulse duration employed in the experiment, the rate
equations are perfectly valid (e.g., see discussion in \cite{LamPRA11,makris09,LamNik13}). The distribution
of the laser intensity within the interaction volume, as determined by the focusing geometry of the laser
beam, has been taken into account. Presumably, the term ``minimal model" has to do with the 
following additional assumptions/approximations adopted by the authors: (a) Only sequential channels were included 
in the rate equations. 
(b) The field (intensity) fluctuations, inherent in  SASE FELs, and specifically FLASH pulses
\cite{Ack07,richter,Mitz08-09}, were not included in the calculations.
(c) All ionic species were assumed to be produced by single-photon ionization.
(d) All of the single-photon cross sections entering the model were assumed to have the same value.

If the purpose of the work in \cite{Guichard13}  was to obtain a fit to the data, it appears that it has
succeeded reasonably well, for the ionic species up to Ne$^{4+}$. However, in order to understand the physical meaning of the fit, we need 
to examine the validity of the underlying assumptions/approximations. Approximation (a) above is
justified, because the direct channels, being of much higher order of non-linearity, are expected to begin
competing with the sequential only for quite short pulses; say below 5 fs, as demonstrated  in
\cite{LamPRA11}. Approximation (b) may be useful in assessing some general features of
the data, but the results may at times be misleading. Approximations (c)
and (d), however, are quite problematic. The difficulty stems from the fact that
the rate equations, which bears repeating are perfectly valid in this context, imply energy conservation
in terms of number of photons absorbed in each transition and the corresponding ionization threshold. This
is a condition inherent in the notion of the cross section. Thus, as discussed in detail below,  
although ions up to Ne$^{4+}$ are produced mainly in a sequence of single-photon
absorptions, from there on it is only two- and three-photon processes that enter the sequence
of allowed transitions.  In the absence of those channels, ionic species beyond Ne$^{4+}$ cannot be
populated through energy conserving processes. Inserting single-photon cross sections for those species
and adjusting parameters may of course produce populations for those species, but the underlying process
is devoid of physical meaning. 

Coming now to the present formulation, the main addition to our earlier paper
\cite{LamPRA11} is the detailed inclusion of all single-photon channels, as well as a careful estimation
of the two- and three-photon cross sections. But perhaps more importantly, accounting for the
field fluctuations, is shown to be of crucial importance in the interpretation of the experimental data. 
As has been shown in great detail in \cite{NikLam12-13}, for a quantitative comparison with
experimental data pertaining to SASE-FEL radiation, it is imperative to include a
stochastic model that describes accurately the statistical properties of the pulses, 
pertaining to the source and the conditions of the experiment. This is an issue
sufficiently important to merit further elaboration at this point. In the absence of intensity
fluctuations, it is basically the laser bandwidth that matters. In that case, for a pulsed source, the
effective bandwidth is determined by the combination of the Fourier and the stochastic bandwidths. But
for a source with intensity fluctuations, the particular statistics underlying the fluctuations are
of crucial importance. Intensity fluctuations entail a spiky temporal structure, under an envelop 
determined by the pulse duration and the usually present monochromator. As a consequence of the
intensity fluctuations, an important parameter is the coherence time, which can be and often has been
determined  through the measurement of a two-photon autocorrelation function \cite{Mitz08-09}. 
In essence, it provides a measure of the duration of the dominant intensity spike under the complicated spiky
structure of a pulse with the stochastic properties of the SASE FEL. There is then an overall (total) bandwidth 
associated with this coherence time which for a Gaussian pulse and Gaussian correlated noise 
is given by \cite{NikLam12-13}
\bea
\Delta\omega = \sqrt{\Delta\omega_{\min}^2+\gamma^2},
\eea
where $\Delta\omega_{\min}$ is the Fourier-limited bandwidth of the pulse and $\gamma = \sqrt{8\pi\ln(2)/T_c}$. 
In general, one cannot account for this bandwidth e.g., by simply correcting the parameters entering the equation of motion for 
the density matrix of the atomic system \cite{NikLam12-13}. That is why detailed modelling of the field fluctuations
is an absolutely necessary ingredient of the rate equations, which as a result are stochastic differential equations\cite{NikLam12-13}.

In the following section 2, we present a detailed discussion of the ionization channels
and the respective orders of non-linearity, included in our calculations. The 1s electrons are assumed
frozen because, for photon energy 93 eV and peak intensities less than $10^{16} $W/cm$^{2}$ the 1s 
electrons are too strongly bound to be affected. The details of the rate equations, including the
interaction volume features are presented in section 3, followed by section 4 in which we present
results without and with field fluctuations, thus demonstrating their significance in interpreting
experimental data. In the last section 5, we summarize our findings as well as certain
issues that remain open. 

    
\section{Ionization paths for Neon at 93eV}
\label{sec2}

\begin{figure}
\begin{center}
\includegraphics*[scale=1.2]{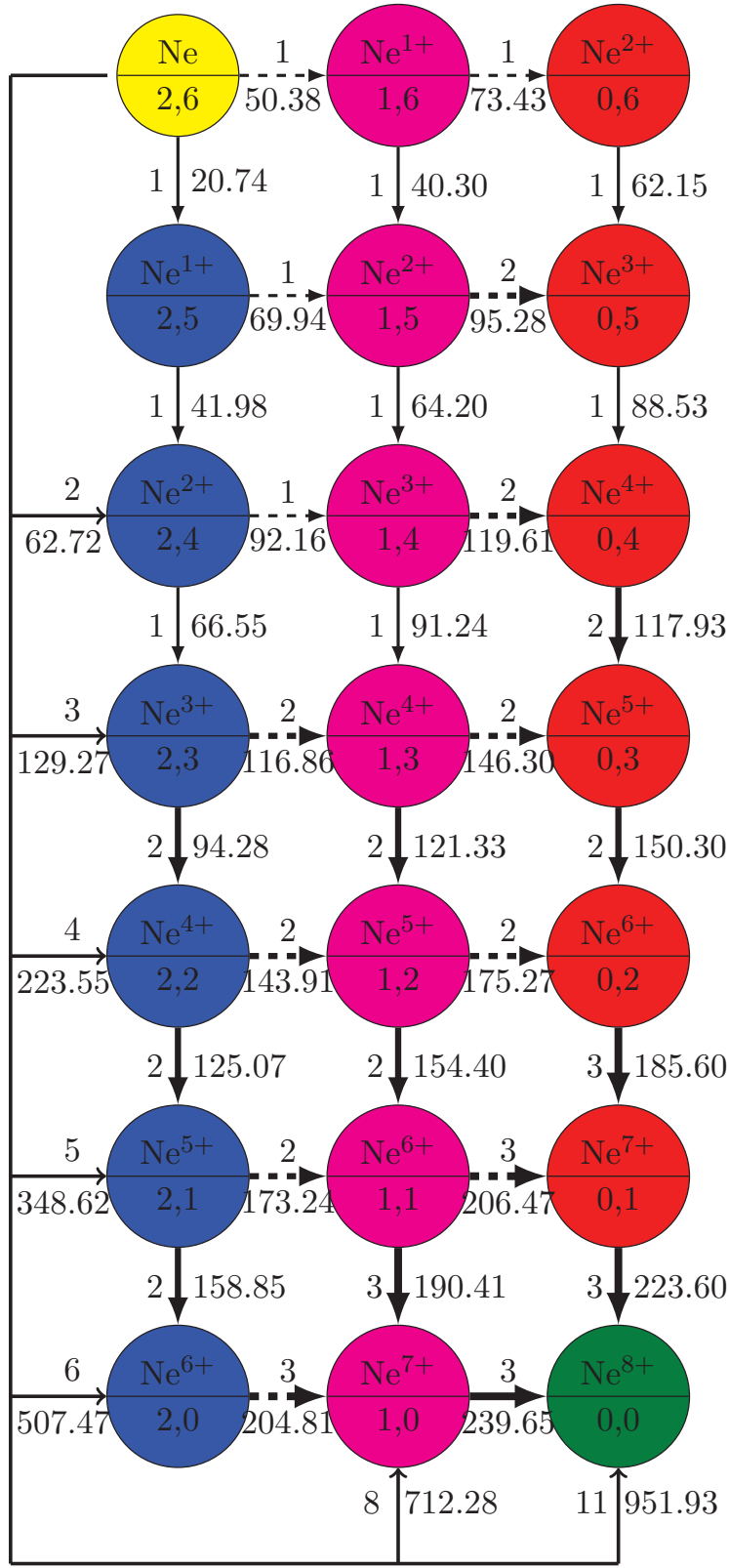}
\caption{\label{diagram_fig}  Ionization paths for Neon. The numbers close to each arrow denote the ionization potential in eV and the number of photons required for ionization assuming photons of 
 energy 93 eV. Different colours (columns) denote ion species with 0, 1 and 2 holes in the 2s shell. }
\end{center}
\end{figure}

A concise summary of all ionization channels included in our calculations is shown in Fig. \ref{diagram_fig}, 
listing important quantities, such as ionization potentials and the order of each transition, together with the flow of charge indicated by arrows. This ``flow chart" includes ejections of electrons from both the 2s and 2p
shells, where each circle corresponds to a particular ionic species with the two numbers in its lower half
denoting the number of remaining electrons in the 2s and the 2p shells.  This is because ionic species at different
internal states appear and disappear during the ionization of neutral Neon at the particular photon energy. 
For example, one can see that Ne$^{3+}$ appears in three different states, namely $\ket{2s^2,2p^3}$, $\ket{2s^1,2p^4}$ and 
$\ket{2s^0,2p^5}$. As will be seen later on, in a sequential ionization process, a given ion at different internal states may 
lead to the same or to different internal states of the next higher ion, through
substantially different cross-sections. For a quantitative comparison to experimental data, the inclusion
of all of these intermediate ionic species is therefore necessary. In the following, the ion Ne$^{q+}$ in the state $\ket{2s^a,2p^b}$ is denoted by Ne$_{a,b}^{q+}$, while charge conservation implies  
\bea
a+b=8-q\equiv \tilde{q}.
\label{charge_conservation}
\eea
In view of Eq. (\ref{charge_conservation}), 
each ionic species is uniquely defined by the two labels $a$ and $b$. 

Careful inspection of the ``flow chart" does reveal some interesting regularities in the underlying
processes, which will help us elucidate the results. More precisely, for the ions up to Ne$^{4+}$
single-photon and two-photon sequential ionization channels co-exist. On the other hand, the ions
Ne$^{5+}$ and Ne$^{6+}$ can be created only through two-photon sequential ionization, whereas three
photons are required for the last two ionic species, namely, Ne$^{7+}$ and Ne$^{8+}$.  
The direct  multiple ionization channels depicted in Fig. \ref{diagram_fig} lead from Ne to Ne$^{q+}$, with 
$q\geq 2$, and pertain to the multiphoton ejection of more than one electrons from the 2p shell of neutral Neon only, 
since  channels that involve  electron ejection from the 2s shell are expected to have considerably smaller cross sections.
 In principle,
when energetically allowed, $n$-photon $m$-electron ejection (with $n\geq m$) can always occur from any
ionic species, and actually it does not necessarily require any electron correlation. As a first approximation, however, 
we have included only direct ionization channels from neutral Neon, since this is mainly present for short times. 
Our simulations, showed that for the parameters of the experiment, the contribution of these channels 
was negligible and thus there was no need for including direct ionization of intermediate ions.

Finally, it should be emphasized that the ``flow chart" of Fig. \ref{diagram_fig} corresponds to photon
energy $93$ eV, with the ionization potentials obtained from the codes in Ref. \cite{cowan}. The counterpart of this 
``flow chart"  for photon energy  $90.5$ eV differs only in the order of the transition 
Ne$_{2,4}^{2+} \to $Ne$_{1,4}^{3+}$, which becomes a two-photon process. In either case, it is the 
ratio of the cross sections for the transitions Ne$_{2,4}^{2+} \to $Ne$_{2,3}^{3+}$ and Ne$_{2,4}^{2+} \to
$Ne$_{1,4}^{3+}$ that determines the dominant ionization path followed by the system. As shown in Table
\ref{tabCS}, the single-photon cross-section for Ne$_{2,4}^{2+} \to $Ne$_{2,3}^{3+}$ at 93 eV is
considerably larger than the corresponding cross-section for Ne$_{2,4}^{2+} \to $Ne$_{1,4}^{3+}$.
Moreover, as discussed later on, the manifold of (near) resonances at 90.5 and 93 eV are very similar,
and thus the relative strength of the cross sections remain practically  the same. Therefore, considering
either of the  photon energies reported in the experiment of \cite{Guichard13}, is not expected to modify
or introduce any prominent distinguishing features in the yields of  Ne$^{2+}$ and Ne$^{3+}$, in agreement with the
reported experimental data. As a result, for the sake of concreteness, our theoretical analysis has been
focused on photons of energy 93 eV. As confirmed later on, our theoretical results for the laser intensity
dependence of the ionic yields are in a very good agreement with those of the experiment for both 90.5 and 93 eV. 

\section{Theory vs Experiment}
\label{sec3}

As discussed above, our model focuses on the populations of the ionic species up to Ne$^{8+}$ and includes different sequential and direct  ionization paths from both of the 2s and 2p shells. 
Although in the experimental data, the highest observable ionic species was Ne$^{6+}$, it is necessary to include the rate equations for 
Ne$^{7+}$ and Ne$^{8+}$, because otherwise the population of Ne$^{6+}$ would increase monotonically, reaching eventually the value 1, violating thus the physical reality of the experiment.
Throughout our simulations, the complete set of differential equations that governs the populations of the ionic species during the pulse is the following: 
\bea
\label{rate_eqs1}
\frac{dN_{2,6}}{dt} &=& -\sigma_{2,6;1,6}^{(1)} F N_{2,6} 
-\sigma_{2,6;2,5}^{(1)} F N_{2,6} \nonumber\\
&&}-\sum_{b=0}^{4}\sigma_{2,6;2,b}^{(6-b)} F^{6-b} N_{2,6}-\sigma_{2,6;1,0}^{(8)} F^{8} N_{2,6}
-\sigma_{2,6;0,0}^{(11)} F^{11} N_{2,6\\
\frac{dN_{2,5}}{dt} &=& \sigma_{2,6;2,5}^{(1)} F N_{2,6} 
-\sigma_{2,5;1,5}^{(1)} F N_{2,5}
-\sigma_{2,5;2,4}^{(1)} F N_{2,5}\\
\frac{dN_{1,6}}{dt} &=& \sigma_{2,6;1,6}^{(1)} F N_{2,6}
-\sigma_{1,6;0,6}^{(1)} F N_{1,6}
-\sigma_{1,6;1,5}^{(1)} F N_{1,6}\\
\frac{dN_{2,4}}{dt} &=& \sigma_{2,5;2,4}^{(1)} F N_{2,5}+\sigma_{2,6;2,4}^{(2)} F^{2} N_{2,6 }\nonumber\\
&&
-\sigma_{2,4;2,3}^{(1)} F N_{2,4}
-\sigma_{2,4;1,4}^{(1)} F N_{2,4} \\
\frac{dN_{1,5}}{dt} &=&\sigma_{2,5;1,5}^{(1)} F N_{2,5}
+\sigma_{1,6;1,5}^{(1)} F N_{1,6}\nonumber\\
&&
-\sigma_{1,5;1,4}^{(1)} F N_{1,5}
-\sigma_{1,5;0,5}^{(2)} F^{2} N_{1,5} \\
\frac{dN_{0,6}}{dt} &=& \sigma_{1,6;0,6}^{(1)} F N_{1,6}
-\sigma_{0,6;0,5}^{(1)} F N_{0,6}\\
\frac{dN_{2,3}}{dt} &=&\sigma_{2,4;2,3}^{(1)} F N_{2,4}+\sigma_{2,6;2,3}^{(3)} F^{3} N_{2,6} \nonumber\\
&&-\sigma_{2,3;2,2}^{(2)} F^2 N_{2,3}
-\sigma_{2,3;1,3}^{(2)} F^2 N_{2,3}\\
\frac{dN_{1,4}}{dt} &=&\sigma_{2,4;1,4}^{(1)} F N_{2,4}
+ \sigma_{1,5;1,4}^{(1)} F N_{1,5}\nonumber\\
&&
-\sigma_{1,4;1,3}^{(1)} F N_{1,4}
-\sigma_{1,4;0,4}^{(2)} F^2 N_{1,4}\\
\frac{dN_{0,5}}{dt} &=& \sigma_{0,6;0,5}^{(1)} F N_{0,6}+\sigma_{1,5;0,5}^{(2)} F^{2} N_{1,5}
-\sigma_{0,5;0,4}^{(1)} F N_{0,5}\\
\frac{dN_{2,2}}{dt} &=& \sigma_{2,3;2,2}^{(2)} F^2 N_{2,3} +\sigma_{2,6;2,2}^{(4)} F^{4} N_{2,6 }\nonumber\\
&&
- \sigma_{2,2;2,1}^{(2)} F^2 N_{2,2} 
-\sigma_{2,2;1,2}^{(2)} F^2 N_{2,2}\\
\frac{dN_{1,3}}{dt} &=&\sigma_{2,3;1,3}^{(2)} F^2 N_{2,3}
+\sigma_{1,4;1,3}^{(1)} F N_{1,4}\nonumber\\
&&
-\sigma_{1,3;1,2}^{(2)} F^2 N_{1,3}
-\sigma_{1,3;0,3}^{(2)} F^2 N_{0,3}\\
\frac{dN_{0,4}}{dt} &=& \sigma_{0,5;0,4}^{(1)} F N_{0,5}
+\sigma_{1,4;0,4}^{(2)} F N_{1,4}
-\sigma_{0,4;0,3}^{(2)} F^2 N_{0,4} \\
\frac{dN_{2,1}}{dt} &=&\sigma_{2,2;2,1}^{(2)} F^2 N_{2,2} +\sigma_{2,6;2,1}^{(5)} F^{5} N_{2,6 }\nonumber\\
&&
- \sigma_{2,1;2,0}^{(2)} F^2 N_{2,1}
- \sigma_{2,1;1,1}^{(2)} F^2 N_{2,1} \\
\frac{dN_{1,2}}{dt} &=&\sigma_{2,2;1,2}^{(2)} F^2 N_{2,2} 
+\sigma_{1,3;1,2}^{(2)} F^2 N_{1,3}\nonumber\\
&&
- \sigma_{1,2;1,1}^{(2)} F^2 N_{1,2}
- \sigma_{1,2;0,2}^{(2)} F^2 N_{1,2}\\
\frac{dN_{0,3}}{dt} &=& \sigma_{0,4;0,3}^{(2)} F^{2} N_{0,4}
+\sigma_{1,3;0,3}^{(2)} F N_{1,3}
-\sigma_{0,3;0,2}^{(2)} F^2 N_{0,3}\\
\frac{dN_{2,0}}{dt} &=&\sigma_{2,1;2,0}^{(2)} F^2 N_{2,1} +\sigma_{2,6;2,0}^{(6)} F^{6} N_{2,6 }
- \sigma_{2,0;1,0}^{(3)} F^3 N_{2,0}\\
\frac{dN_{1,1}}{dt} &=&\sigma_{2,1;1,1}^{(2)} F^2 N_{2,1} 
+\sigma_{1,2;1,1}^{(2)} F^2 N_{1,2} \nonumber\\
&&
- \sigma_{1,1;1,0}^{(3)} F^3 N_{1,1}
- \sigma_{1,1;0,1}^{(3)} F^3 N_{1,1} \\
\frac{dN_{0,2}}{dt} &=& \sigma_{0,3;0,2}^{(2)} F^{2} N_{0,3}
+\sigma_{1,2;0,2}^{(2)} F N_{1,2}
-\sigma_{0,2;0,1}^{(3)} F^3 N_{0,2}\\
\frac{dN_{1,0}}{dt} &=&\sigma_{2,0;1,0}^{(3)} F^{3} N_{2,0}+\sigma_{2,6;1,0}^{(8)} F^{8} N_{2,6 }\nonumber\\
&&
+\sigma_{1,1;1,0}^{(3)} F^{3} N_{1,1}
-\sigma_{1,0;0,0}^{(3)} F^3 N_{1,0}\\
\frac{dN_{0,1}}{dt} &=& \sigma_{0,2;0,1}^{(3)} F^{3} N_{0,2}
+\sigma_{1,1;0,1}^{(3)} F N_{1,1}
-\sigma_{0,1;0,0}^{(3)} F^3 N_{0,1}\\
\frac{dN_{0,0}}{dt} &=& \sigma_{1,0;0,0}^{(3)} F^3 N_{1,0}+\sigma_{0,1;0,0}^{(3)} F^3 N_{0,1} +\sigma_{2,6;0,0}^{(11)} F^{11} N_{2,6 }
.
\label{rate_eqs8}
\eea
Here, $N_{a,b}$ refers to the population of the ionic species Ne$_{a,b}^{q+}$. 
A term like 
$\sigma_{a,b;a^\prime,b^\prime}^{(n)} F^{n} N_{a^\prime,b^\prime}$ represents an $n$-photon process leading from 
Ne$_{a,b}^{q+}$ to Ne$_{a^\prime,b^\prime}^{q^\prime+}$ with the corresponding (generalized) cross-section  $\sigma_{a,b;a^\prime,b^\prime}^{(n)}$ in units of cm$^{2n}$sec$^{n-1}$, while $F(t)$ is the time-dependent photon flux in photons/cm$^2$sec. 
The total ionization yield for Ne$^{q+}$ is obtained by summing up the yields at all possible states of Ne$^{q+}$ i.e., 
\bea
N_q = \sum_{a} N_{a,8-q-a}, 
\eea
with $\max\{0,2-q\} \leq a\leq \min\{2,8-q\}$.

In setting up the rate equations, all single-photon cross sections were obtained from \cite{cowan}. The values of the two- and three-photon cross sections were obtained through a  combination of scaling \cite{scaling} with proper accounting for the level structure of the respective species, in order to identify the possible influence of near-resonant intermediate states. The final set
of cross sections employed in the calculations for photon energy at 93 eV is listed in Table \ref{tabCS}. The reader familiar with the range of magnitudes of multiphoton ionization cross sections may notice that some of the two-photon and three-photon cross sections tend to be in the range of the larger values. This is due to the presence of intermediate near resonances some of which are given in Table \ref{tabCS2}. It should be emphasized here that the landscape of (near) resonances we have  found by means of  \cite{cowan} at 93 eV, remains practically the same at 90.5 eV. Hence, the above set of rate
equations and the associated cross-sections listed in table \ref{tabCS} are expected to describe equally
well the ionization of Neon at both photon energies. 

The closed set of rate equations (\ref{rate_eqs1})-(\ref{rate_eqs8})  is to be solved under a realistic pulse, specified by the conditions in the experiment under consideration. For
quantitative comparison with experimental data, however, one must take into account the spatial
distribution of the laser intensity within the interaction volume. Throughout our simulations we have
considered a Gaussian beam whose intensity is given by  
\bea
I(r,z;t) = I(t)\frac{w_0^2}{w(z)^2} \exp\left [ -\frac{2r^2}{w(z)^2}\right]\equiv I(t) I(r,z)
\label{gauss1}
\eea
where $w(z) = w_0\sqrt{1+(z/z_R)^2}$ is the beam radius, $z_R$ is the Rayleigh length and $w_0$ is the beam waist. The temporal profile of the intensity is also taken to be a Gaussian of the form   
 \bea
I(t) = I_0\exp\left [-\frac{(t-t_0)^2}{\tau^2}\right ], 
\label{gauss2}
\eea 
where $I_0$ is the peak intensity, $t=t_0$ is the center of the pulse, and $\tau$ is the pulse duration.  
The FWHM of $I(t)$ is given by $2\sqrt{\ln(2)}\tau$. 

The solution of the rate equations takes place on a three-dimensional grid for $(r,z;t)$, with 
$r\in[0,R_{\max}]$, $z\in[-Z_{\min},Z_{\max} ]$ and $t\in[0,T_f]$. For a given peak intensity $I_0$, 
we choose a point in space $(r,z)$ and the equations are solved numerically in time from $t=0$ to $t=T_f$, with the initial condition $N_{2,6} = 1$ and $N_{a,b}=0$, otherwise. The flux entering the equations is given by 
\bea
F(r,z;t) = \frac{0.624\times 10^{19}}{\hbar\omega_{ph}} I(r,z;t)
\eea
where the photon energy $\hbar\omega_{ph}$ is given in eV and the intensity is in W/cm$^{2}$. 
The same procedure is followed for the 
next spatial point, until the entire spatial grid is covered. In this way, one obtains the yields for the values  of the intensity at all  points on the spatial grid. Finally,  performing a numerical spatial integration one  obtains the total ionic yields at the chosen peak intensity $I_0$. 

The resulting theoretical predictions, at this point for a Gaussian Fourier-limited pulse, together with the experimental yields reported in \cite{Guichard13}  are presented in Fig. \ref{Yields1_fig}, for three different pulse durations.  All of the parameters entering Eqs. (\ref{gauss1}) and (\ref{gauss2}) are in agreement with
the ones adopted in \cite{Guichard13}. There is an overall very good agreement with the experimental data
for all species, especially for pulses with FWHM 15fs and 20fs. There is a disagreement for Ne$^{1+}$ at
high intensities, as well as for the ions above  Ne$^{3+}$, which were also present in the comparison of
the experimental data with the ``minimal model" of  Ref. \cite{Guichard13}.  The present theoretical model,
however, produces much better agreement with the experimental data for all ions.  Although in some cases
the observed discrepancies are within the reported experimental uncertainties,  the question arises as to
whether  a better agreement could be obtained in the framework of the present theoretical model under the constraint of the experimental conditions reported in \cite{Guichard13}. 
Having performed extensive simulations,  we have reached the conclusion that no dramatic
improvement can be expected with Fourier-limited temporal pulse shapes.

\begin{figure}
\begin{center}
\includegraphics*[scale=0.8]{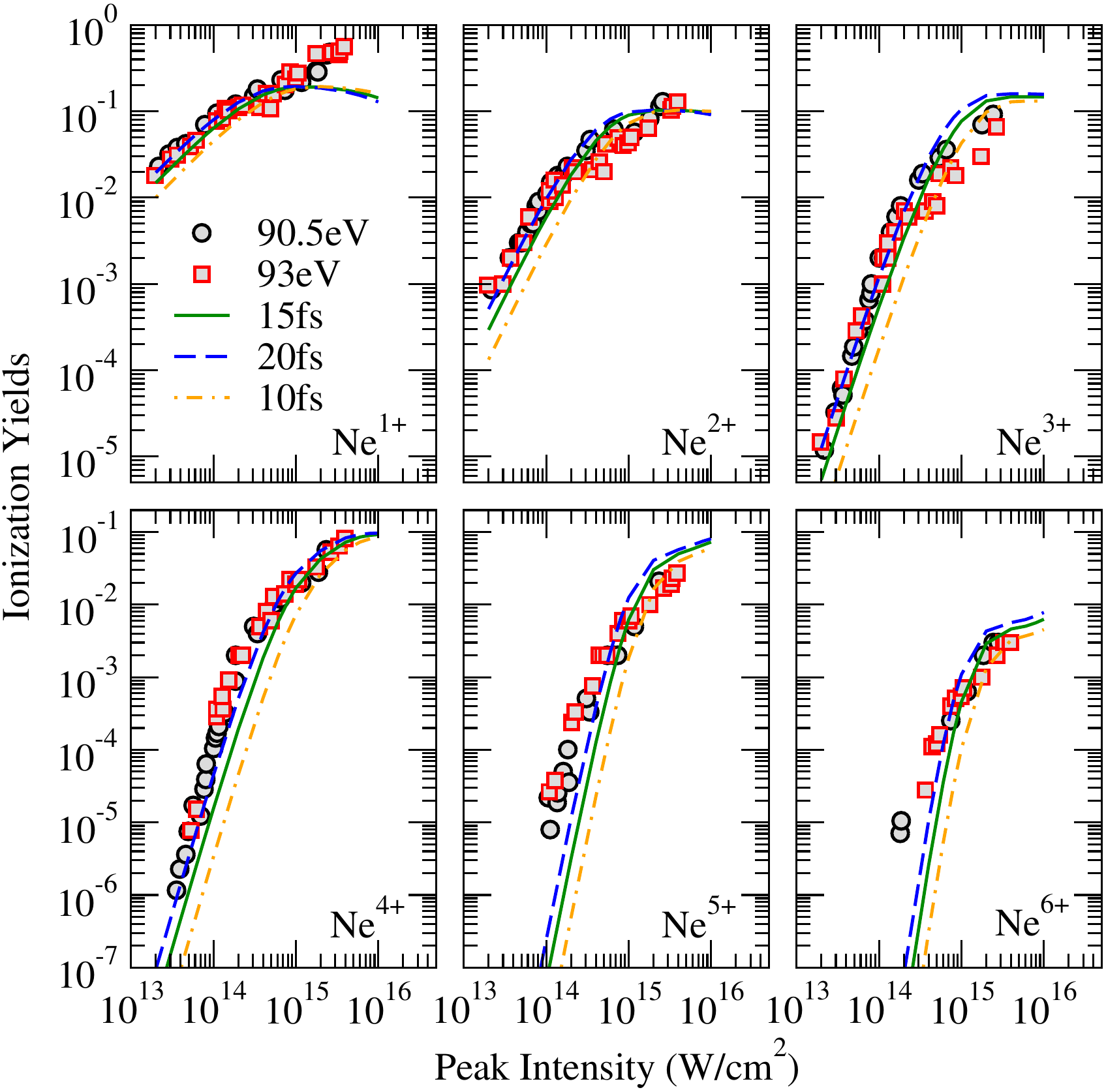}
\caption{\label{Yields1_fig} Experiment vs theory: Ionization yields for Neon. The symbols are experimental data for 
93  and 90.5 eV, and have been extracted from \cite{Guichard13}. The curves are theoretical results after spatial integration for Fourier-limited pulses of various durations (FWHM). }
\end{center}
\end{figure}

To those familiar with the properties of SASE-FEL radiation, this would not come as a surprise. Because 
an important feature of SASE-FEL radiation, which has not been included in our model so far,  is the
presence of stochastic intensity  fluctuations which are well represented by Gaussian-correlated noise
\cite{Ack07,richter,Mitz08-09}. 
As a consequence a pulse of duration $\tau$ typically exhibits a number of narrower spikes whose duration
is basically associated with the so-called coherence time $T_c$. In fact, the coherence time at FLASH 
at the time of the experiments under consideration was estimated to be about $6$ fs, which means that for 
pulse durations 15$\pm$5 fs, roughly speaking there were a few spikes per pulse \cite{Ack07,richter,Mitz08-09}.  We have developed a
rather efficient numerical approach for the simulation of such pulses, which captures all of the
essential features for  the problem under consideration. For example, our approach was
recently applied to the study of the line-shape of an Auger resonance in Kr \cite{NikLam12-13}, 
in excellent agreement with related  experimental observations at FLASH \cite{MazJPB12}.  Our numerical
approach, as described in detail elsewhere \cite{NikLam12-13}, has been employed in this work as well, in 
order to address the aforementioned persistent discrepancy between theory and experiment, for 
Fourier-limited pulses.  

In taking this step, the only thing that changes in our model is the temporal profile of the pulse
entering our equations, which is no longer given by Eq. (\ref{gauss2}), but is instead a stochastic
function generated numerically according to the algorithm of  \cite{NikLam12-13}. Under the stochastically
fluctuating pulses, the equations must now be solved for a number of randomly generated pulses, with the
quantities of interest being the ionic yields averaged over an appropriate number of fluctuating pulses 
which, by the way, is exactly what is done in the collection of the experimental data. In the case under
consideration, the experimental data presented in  \cite{Guichard13} were averaged over 500 consecutive
FEL pulses, and we have  thus chosen the same number in our simulations.   As shown in Fig.
\ref{Yields2_fig}, we obtain excellent agreement with the experimental data for all ionic species, including Ne$^{5+}$ and Ne$^{6+}$. The
only remaining discernible discrepancy appears for Ne$^{1+}$ at the higher peak intensities, which is also
the case in \cite{Guichard13}. Possible reasons for this discrepancy are discussed in the section that follows.

\begin{figure}
\begin{center}
\includegraphics*[scale=0.8]{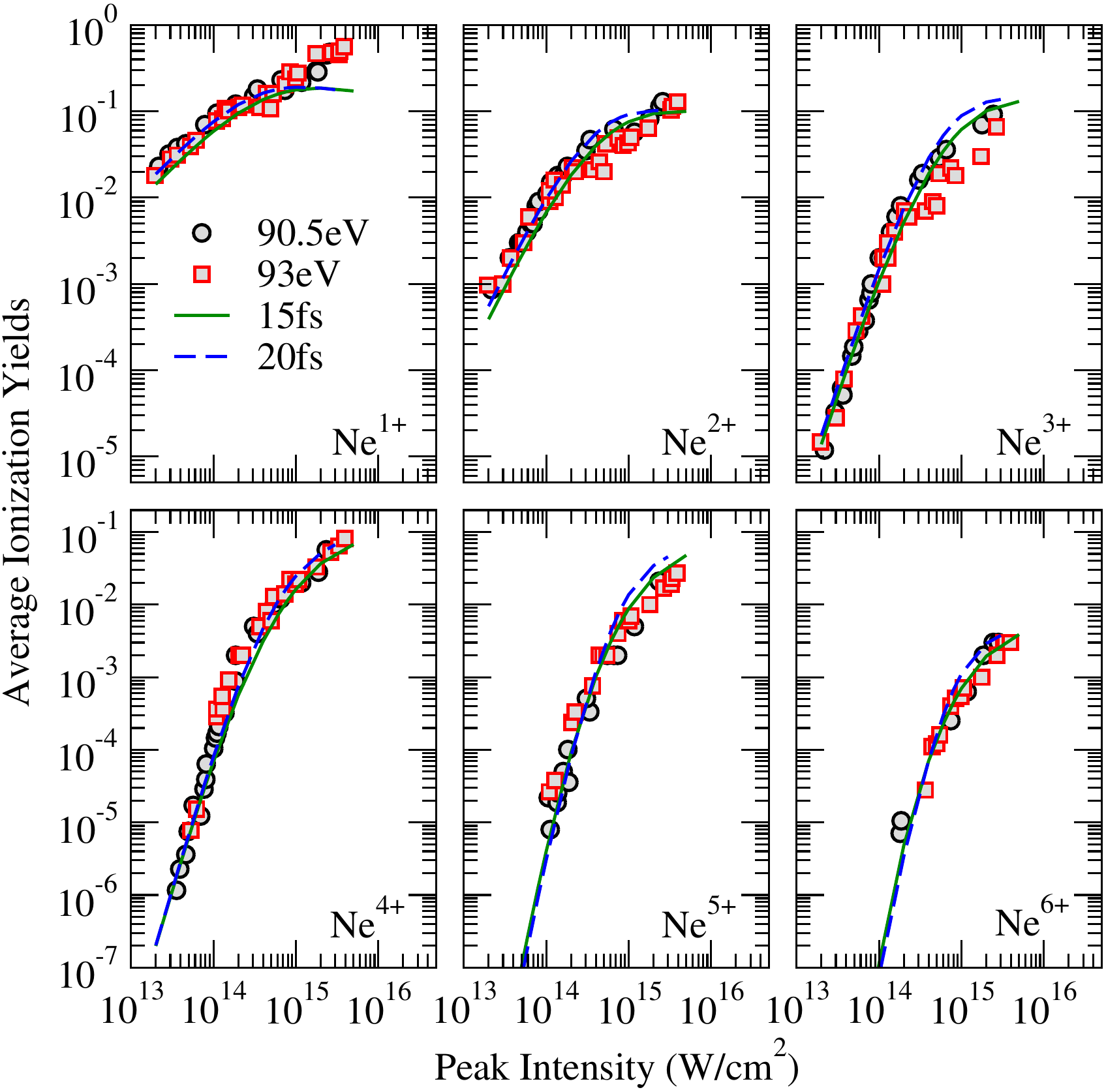}
\caption{\label{Yields2_fig} Experiment vs theory: Ionization yields for Neon. The symbols are experimental data for 93  and 90.5 eV, and have been extracted from \cite{Guichard13}. The curves are theoretical results after spatial integration, averaged over 500 fluctuating pulses with $T_c = 6$ fs and two different 
pulse durations (FWHM). }
\end{center}
\end{figure}


\section{Concluding remarks and discussion}

Motivated by the recent work of Guichard et al. \cite{Guichard13}, we have extended our previous work
on the ionization of the Neon at photon energies $\sim 93$ eV, in the framework of Lowest non-vanishing
Order Perturbation Theory (LOPT).  Aiming at an approach that can serve as an example for the
interpretation of similar experiments under FEL radiation, we have endeavoured to include in the theory as much
quantitative detail as possible. In that sense, our objective has been to develop an approach
with predictive as well as interpretative capability. The excellent agreement between theory and
experiment in the laser intensity dependence of all available ionic yields, that we have obtained,
confirms the quantitative validity of LOPT \cite{makris09} and the associated rate equations, 
incorporating sequential and if necessary direct multiple ionization channels, in the context of experiments under present day FEL sources. To this end, however, all possible atomic shells involved in the ionization have to be taken into account, with the assignment of reasonable cross sections that are consistent with the
atomic structure. But that may not be enough. As we have demonstrated in this work, under SASE-FEL sources
and particularly in the presence of  non-linear processes, the realistic and accurate inclusion of the 
stochastic properties of the source, accounting for the intensity fluctuations, is mandatory for the
quantitative interpretation and understanding of experimental data on multiple ionization.

We have thus grounds to argue that a quantitative interpretation of experimental data, such as those
considered in this paper, which is consistent with the underlying physics, is best served by a
sufficiently elaborate theoretical model. We invite the reader to contrast our fit of the higher ionic
species, namely Ne$^{5+}$ and Ne$^{6+}$, with that in \cite{Guichard13}. The inadequacy of the minimal
model for those ions is traceable directly to the absence of the non-linear processes beyond Ne$^{4+}$
which are necessary for the generation of the higher ions.
    
A discrepancy between theory and experiment for Ne$^{1+}$ at the higher intensity range has persisted
through all of our calculations, as is also the case with the minimal model results \cite{Guichard13}.
Since the production of Ne$^{1+}$ involves only single photon processes, with known cross sections and 
without any complications of atomic structure, this discrepancy is somewhat puzzling. One possible cause
of this behaviour may have to do with the way data at different intensities were taken in the 
measurements. Usually, the intensity is decreased by inserting a linear absorber between the source and
the gas chamber. In that case, the geometry of the interaction volume is the same for all intensities, 
and this is how it has been modelled in our theory. According to
the experimental papers, however, the intensity has been controlled by moving the focal spot of the
radiation from the center of the atomic beam. As we understand it, this may indeed expose most of the
atoms in the gas jet to lower intensities, but at the same time the geometry of the interaction volume is
altered. Having encountered this issue in much earlier work \cite{makris09} on data by the same group, we
have noted that some sort of volume ``renormalization" must have been involved. Without access to the 
details of this procedure, there is not much we can add by way of answer, other than leaving the above
comments as a conjecture.

There is one additional aspect that is worth discussing in brief before closing. It has to do with the 
possibility that ionization plus excitation, often called also ``shake up", may leave some ions in an
excited state, which can then be ionized in the next step. Let us, for the sake of clarity, take a 
specific example. Under photon energy of 93 eV, in addition to ejecting one 2p electron from neutral Neon, there is a 
single-photon two-electron process which leads to Ne$^{1+}$, with one electron for example in the 3d state.
This process can be repeated in the next step, giving rise to an analogous excited state of Ne$^{2+}$. 
In the next step, this two-electron process is energetically forbidden. The electrons in the created
excited states can be ionized by single-photon absorption. Although, we have not been able to find
experimental data or theoretical calculations for these two-electron processes in Neon, from information 
for other atoms and molecules \cite{MazJPB12,WijKel87,Kos87,SanMar91,NakNik02}, 
we do know that typically the cross section for ionization plus excitation
amounts to a few percent, say 2-4\%, of the dominant single-photon ionization cross section of a 2p
electron, which leaves the ion in the ground state. 
That is because such single-photon two-electron
processes rely on electron correlation which is weak compared to the ejection of one electron. For pulse durations in the range of 10-50 fs, those excited states do not have the time to undergo spontaneous
emission. Knowing the ionization cross sections of such excited states, which are in the range of
$10^{-20}$ to $10^{-19}$cm$^{2}$, we can estimate rather reliably the amount of Ne$^{4+}$ that may survive
in the excited state. It is the excited state of this species that is of interest here, because the
single-photon ionization of the excited electron would compete with the two-photon ionization of the ground state. Given that only a very small amount, at best no more than 2\% or so, may be present during
the pulse, we can calculate the laser intensity at which its ionization would compete with the two-photon
process of the ground state. The result is that for intensities larger than $10^{13}$ W/cm$^{2}$, the
contribution of that single photon process is insignificant, in comparison to the two-photon one, even if
the two-photon cross section is quite small. We should perhaps remind the reader that the rate of a
single-photon process, which depends linearly, can compete with a two-photon process, which depends
quadratically on the photon flux, only at lower intensities. It is nevertheless conceivable that a minute
trace of those shake-up processes may be present in the data. Their identification would, however, require
data well beyond the detection of ionic yields.

But even within the limits of the detection of ionic yields, there are interesting details which
illustrate the interplay of various non-linear processes. One such example is shown in Fig. \ref{Yields3_fig}, 
where we have plotted ionic yields without interaction volume averaging for a Fourier-limited pulse of 15 fs. 
As the reader will notice, beginning with Ne$^{4+}$, the yields for the higher ionic species exhibit dips of increasing depth,  at the higher 
laser peak intensities.  
Although the
volume averaging is an experimental reality, still the detection of ionic yields, at the center of the
interaction volume where the intensity is constant, is in principle possible, but much more demanding
experimentally. Nevertheless having that behaviour at the theoretical level, is useful in that it does
provide a point of comparison of different theoretical approaches, while at the same time illustrating details of the dynamics 
of the system.  We do realize, as we have tested it, that the presence of fluctuations will most likely smooth out the dips appearing at relatively high intensities. On the other hand, one of the directions pursued in the continuing development of FEL sources is 
to reach practically Fourier-limited pulses. Thus, our example albeit idealized at this point in time, may be useful as a point of calibration 
in the near future. 
 
\begin{figure}
\begin{center}
\includegraphics*[scale=0.6]{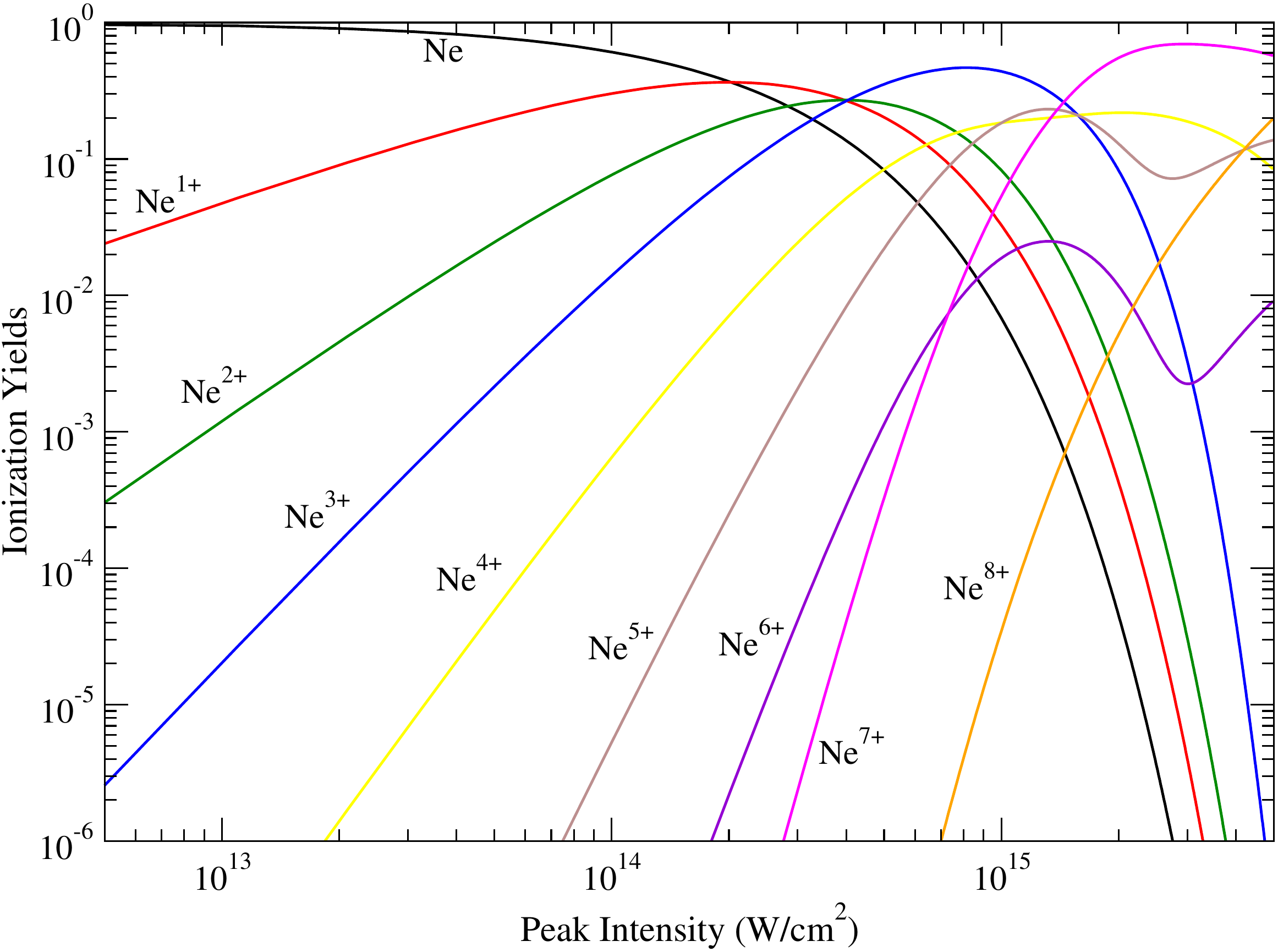}
\caption{\label{Yields3_fig} Ionization yields for Neon at 93 eV, for Fourier-limited pulse of FWHM 15 fs, and with no volume expansion effects.  }
\end{center}
\end{figure}
 
\section{Acknowledgments}
The authors wish to acknowledge informative communications with the authors of \cite{Guichard13} concerning the assumptions adopted in their model.This work was supported in part by the European COST Action CM1204 (XLIC).

\appendix

\section{Cross sections}

Figures \ref{Yields1_fig} and \ref{Yields2_fig} have been produced using the cross-sections given in table  \ref{tabCS}. The single-photon cross sections entering the above equations have been obtained by means of the codes in \cite{cowan}, whereas the multiphoton cross-sections have been obtained as explained in section  \ref{sec3} . 

\begin{table}
\caption{\label{tabCS} Values of the cross sections used in obtaining the figures of this work $^{(\rm a)}$.}
\footnotesize\rm
\begin{tabular*}{\textwidth}{@{}l*{15} {@{\extracolsep{0pt plus12pt} } c} }
\br
Cross section $\sigma_{a,b;a^\prime,b^\prime}^{(n)}$& Value (cm$^{2n}$sec$^{n-1}$)  \\
\mr
$\sigma_{2,6;2,5}^{(1)}$ & 4.08$\times10^{-18}$\\ 
$\sigma_{2,6;1,6}^{(1)}$ & 0.57$\times10^{-18}$\\ 
$\sigma_{2,5;2,4}^{(1)}$ & 4.05$\times10^{-18}$\\ 
$\sigma_{2,5;1,5}^{(1)}$ & 0.66$\times10^{-18}$\\ 
$\sigma_{1,6;1,5}^{(1)}$ & 4.51$\times10^{-18}$\\ 
$\sigma_{1,6;0,6}^{(1)}$ & 0.32$\times10^{-18}$\\
$\sigma_{2,4;2,3}^{(1)}$ & 3.97$\times10^{-18}$\\ 
$\sigma_{2,4;1,4}^{(1)}$ & 0.77$\times10^{-18}$\\ 
$\sigma_{1,5;1,4}^{(1)}$ & 4.57$\times10^{-18}$\\ 
$\sigma_{0,6;0,5}^{(1)}$ & 5.06$\times10^{-18}$\\ 
$\sigma_{1,4;1,3}^{(1)}$ & 4.54$\times10^{-18}$\\
$\sigma_{0,5;0,4}^{(1)}$ & 5.20$\times10^{-18}$\\ 
$\sigma_{1,5;0,5}^{(2)}$ & 1.00$\times10^{-52}$\\ 
$\sigma_{2,3;2,2}^{(2)}$ & 5.00$\times10^{-51}$\\ 
$\sigma_{2,3;1,3}^{(2)}$  & 1.00$\times10^{-50}$\\ 
$\sigma_{1,4;0,4}^{(2)}$  & 5.00$\times10^{-51}$\\ 
$\sigma_{2,2;2,1}^{(2)}$  & 1.00$\times 10^{-52}$\\ 
$\sigma_{2,2;1,2}^{(2)}$  & 1.00$\times10^{-51}$\\ 
$\sigma_{1,3;1,2}^{(2)}$  & 1.00$\times10^{-51}$\\ 
$\sigma_{1,3;0,3}^{(2)}$  & 1.00$\times 10^{-49}$\\ 
 $\sigma_{0,4;0,3}^{(2)}$  & 1.00$\times 10^{-51}$\\ 
 $\sigma_{2,1;2,0}^{(2)}$  & 1.00$\times10^{-52}$\\ 
$\sigma_{2,1;1,1}^{(2)}$ & 1.00$\times 10^{-52}$\\ 
$\sigma_{1,2;1,1}^{(2)}$ & 1.00$\times10^{-52}$\\ 
$\sigma_{1,2;0,2}^{(2)}$ & 1.00$\times 10^{-52}$\\ 
 $\sigma_{0,3;0,2}^{(2)}$ & 2.20$\times 10^{-50}$\\ 
$\sigma_{2,0;1,0}^{(3)}$ & 1.00$\times10^{-84}$\\ 
$\sigma_{1,1;1,0}^{(3)}$ & 1.00$\times 10^{-84}$\\ 
$\sigma_{1,1;0,1}^{(3)}$ & 1.00$\times10^{-84}$\\ 
$\sigma_{0,2;0,1}^{(3)}$ & 5.00$\times10^{-81}$\\ 
$\sigma_{1,0;0,0}^{(3)}$ & 1.00$\times 10^{-84}$\\  
$\sigma_{0,1;0,0}^{(3)}$ & 1.00$\times 10^{-84}$\\ 
$\sigma_{2,6;2,4}^{(2)}$ & 1.00$\times10^{-52}$\\ 
$\sigma_{2,6;2,3}^{(3)}$ & 1.00$\times 10^{-84}$\\ 
$\sigma_{2,6;2,2}^{(4)}$ & 3.00$\times10^{-117}$\\ 
$\sigma_{2,6;2,1}^{(5)}$ & 6.00$\times10^{-150}$\\ 
$\sigma_{2,6;2,0}^{(6)}$ & 5.00$\times 10^{-183}$\\  
$\sigma_{2,6;1,0}^{(8)}$ & 4.00$\times 10^{-249}$\\ 
$\sigma_{2,6;0,0}^{(11)}$ & 1.00$\times 10^{-348}$\\ 
 
\br
\end{tabular*}
$^{(\rm a)}$ Single-photon cross-sections have been obtained from \cite{cowan}.
\end{table}

\begin{table}
\caption{\label{tabCS2} Some near-resonant transitions for ionization of Neon at 93 eV as obtained from \cite{cowan}. Note that this table is not by any means exhaustive.}
\footnotesize\rm
\begin{tabular*}{\textwidth}{@{}l*{15} {@{\extracolsep{0pt plus12pt} } c} }
\br
Ion & Transition & Energy (eV)\\
\mr
Ne$^{3+}$ & (2s$^2$ 2p$^3$)$^4$S$_{3/2}$ $\to$ (2s$^2$ 2p$^2$ 7d$^1$)$^4$P$_{1/2}$ & 9.2896E+01\\
Ne$^{3+}$ & (2s$^2$ 2p$^3$)$^2$P$_{3/2}$ $\to$ (2s$^1$ 2p$^3$ 3p$^1$)$^2$P$_{1/2}$ & 9.3092E+01\\
Ne$^{3+}$ & (2s$^1$ 2p$^4$)$^4$P$_{3/2}$ $\to$ (2p$^4$ 3p$^1$)$^4$D$_{5/2}$ & 9.2992E+01\\

Ne$^{4+}$ & (2s$^1$ 2p$^3$)$^1$P$_1$ $\to$ (2s$^1$ 2p$^2$ 4s$^1$)$^1$P$_1$ & 9.2978E+01\\
Ne$^{4+}$ & (2s$^2$ 2p$^2$)$^3$P$_1$ $\to$ (2s$^1$ 2p$^2$ 3p$^1$ )$^3$D$_2$ & 9.2895E+01\\
Ne$^{4+}$ & (2s$^1$ 2p$^3$)$^3$S$_1$ $\to$ (2p$^3$ 3p$^1$ )$^3$P$_2$ & 9.2836E+01\\
Ne$^{4+}$ & (2p$^4$)$^3$P$_0$ $\to$ (2p$^3$ 4s$^1$ )$^3$D$_1$ & 9.3176E+01\\
 
Ne$^{5+}$ & (2p$^3$)$^2$P$_{1/2}$ $\to$ (2p$^2$ 3d$^1$ )$^2$P$_{1/2}$ & 9.3244E+01\\

Ne$^{6+}$ &(2p$^2$)$^3$P$_2$ $\to$  (2p$^1$ 20f$^1$ )$^3$F$_3$ & 1.8598E+02 $^{\rm (a)}$\\
Ne$^{6+}$ &(2p$^2$)$^3$P$_2$ $\to$  (2p$^1$ 20f$^1$)$^3$F$_4$ & 1.8597E+02 $^{\rm (a)}$\\

\br
\end{tabular*}
$^{\rm (a)}$ Two photon transitions.
\end{table}

\end{document}